\documentclass[twocolumn,showpacs,preprintnumbers,amsmath,amssymb,prb,aps]{revtex4-1}

\usepackage{epsfig} 
\usepackage{graphicx}
\usepackage{dcolumn} 
\usepackage{bm}

\makeatletter

\newcommand{\Rmnum}[1]{\expandafter\@slowromancap\romannumeral #1@}
\makeatother

\begin{document}

\title{Investigation of the electronic and thermoelectric properties of Fe$_{2}$ScX (X = P, As and Sb) full Heusler alloys by using first principles calculations}
\author{Sonu Sharma and Sudhir K. Pandey}
\address{School of Engineering, Indian Institute of Technology Mandi, Kamand - 175005, India}
\address{Electronic mail: sonusharma@iitmandi.ac.in}

\date{\today}

\begin{abstract}
By using \textit{ab initio} electronic structure calculations here we report the three new full Heusler alloys which are possessing very good thermoelectric behavior and expected to be synthesized in the laboratories. These are Fe$_{2}$ScP, Fe$_{2}$ScAs and Fe$_{2}$ScSb compound. First two compounds are indirect band gap semiconductors and the last one shows semimetallic ground state. The value of band gap of Fe$_{2}$ScP and Fe$_{2}$ScAs is 0.3 and 0.09 eV, respectively. These compounds show the presence of flat conduction bands along $\Gamma$ - X-direction suggesting for the large electron like effective mass and promising for very good thermoelectric behavior of the compounds. At 200 K, the Seebeck coefficients of Fe$_{2}$ScP, Fe$_{2}$ScAs and Fe$_{2}$ScSb compounds are -770, -386 and -192$\mu$V/K, respectively. The maximum power factor ($PF$) is expected for the \textit{n}-type doping in these materials. The heavily doped Fe$_{2}$ScP and Fe$_{2}$ScAs have $PF$ \textgreater 60 for a wide temperature range, which is comparable to the PF of Bi$_{2}$Te$_{3}$ - a well known and one of the best commercially used thermoelectric material. Present work suggests the possible thermoelectric applicability of these materials in a wide temperature range.

\end{abstract}

\maketitle

\section{Introduction}

In recent years, the demand for the development of clean and renewable energy sources is very urgent, because the natural resources are diminishing and their burning contributes to the global warming. Thermoelectric materials have attracted much attention because of their capability of converting heat directly into electricity and vice versa\cite{tan,wang,ypei,akasaka}.
The efficiency of thermoelectric devices is determined by the dimensionless figure-of-merit ($ZT$), which depends on transport properties of the material such as the Seebeck coefficient ($S$), the electrical conductivity ($\sigma$) and thermal conductivity ($\kappa$ = $\kappa_{e}$+$\kappa_{l}$) and given by\cite{pei,lalonde}

\begin{equation}
{ZT} = S^{2}\sigma T/\kappa
\end{equation}
This equation suggests that, a high $ZT$ requires a large a large $S$ and $\sigma$ and a low $\kappa$. $S$ for a metal or degenerate semiconductors is related with the effective mass ($m^{*}$) by the following formula\cite{synder}:
\begin{equation}
 S = (8\Pi ^{2}k^{2}_{B} /3eh^{2})m^{*}T (\Pi/3n)^{2/3} 
\end{equation}
where Boltzmann constant, $k_{B}$ and electronic charge, $e$ are constants and hence at fixed temperature $T$, $S$ depends on $n$ and $m^{*}$. This equation implies that the $S$ will be more in the materials whose $m^{*}$ is large. 

Most of the Heusler alloys have half metallic ground state with $\sim$100 \% spin-polarization at the Fermi level ($E_{F}$) where majority and minority spin channels usually show metallic and insulating behaviors, respectively. The minority spin channel of these compounds normally shows the presence of flat conduction band (CB) along $\Gamma$ - X direction\cite{sonu, ssonu, yabuuchi, barth, tgraf, jbarth, picozzi}. The presence of flat CB appears to be the generic property of the full Heusler alloys. This flat band will give rise to the large value of $m^{*}$ and  hence large value for $S$ corresponding to minority spin channel. Recently, we have explored this aspect for Co$_{2}$MnGe half-metallic full Heusler alloy. Our results also show the presence of flat CB along $\Gamma$ - X-direction in the spin down channel. At room temperature we got the calculated value of $S$ $\sim$ -550 $\mu$V/K corresponds to the spin-down channel. This value was about 55 times more than that of the spin-up channel. However, the overall value of $S$ is found to be about 10 $\mu$V/K due to the presence of $\sim$1000 times larger $\sigma/\tau$ for up spin channel in comparison to down spin channel\cite{sonu}. This result clearly suggests that if the ground state of the full Heusler alloys happens to be semiconducting then these alloys would have been a very good thermoelectric materials. 

The presence of semiconducting ground state for full Heusler alloys is very rare \cite{meinert}. Interestingly, Yabuuchi et al.\cite{yabuuchi} have predicted the presence of semiconducting ground state having very large $S$ for Fe$_{2}$TiSn and Fe$_{2}$TiSi compounds. The presence of semiconducting ground state also demands nonmagnetic ground state because of the Slater-Pauling rule. For full-Heusler alloys this rule is given by the following equation\cite{galanakis, graf} 
\begin{equation}
m= N_{V}-24
\end{equation}
where $N_{V}$ is valence electron number per formula unit. It is clear from this formula that the magnetic moment vanishes for the compounds having valence electron number equal to 24 and these compounds are semimetallic or semiconducting\cite{galanakis, yabuuchi}.

In the search of the semiconducting full Heusler alloys we come across with a new set of compounds viz. Fe$_{2}$ScP, Fe$_{2}$ScAs and Fe$_{2}$ScSb. These compounds have valence electron number equal to 24 and thus their total magnetic moment is expected to vanish as per Slater-Pauling rule. In the present work we study the electronic and transport properties of these compounds by using full-potential linearized augmented plane wave method based on density functional theory along with the Boltzmann transport theory. First two compounds are found to be indirect band gap semiconductors whereas the last one have semimetallic ground state. As expected these compounds show almost flat conduction band along $\Gamma$ - X-direction. The indirect band gap of about 0.3 and 0.09 eV are found for  Fe$_{2}$ScP and Fe$_{2}$ScAs compounds. The undoped compounds show a very large and negative value of $S$, which suggests that these Heusler alloys are \textit{n}-type thermoelectric materials. Also the power factor ($PF$) versus chemical potential ($\mu$) plots give the qualitative information about the type of doping and suggests that the \textit{n}-type doping can further enhance the $S$ of these compounds. When Fe$_{2}$ScP and Fe$_{2}$ScAs were doped heavily with electrons the $PF$ of these materials found to be closer to the $PF$ of Bi$_{2}$Te$_{3}$. These materials offer good thermoelectric properties for the wide temperature range. Further the calculations of the formation energy of these materials also suggest the possibility of synthesizing them in the laboratories.

\section{Computational details}
By combining the first principles calculations with the BoltzTraP code we have investigated the electronic and thermoelectric properties of new Heusler alloys viz. Fe$_{2}$ScP, Fe$_{2}$ScAs and Fe$_{2}$ScSb. The electronic properties have been investigated by using the full-potential linearized augmented plane-wave (FP-LAPW) method within the density functional theory (DFT) implemented in WIEN2k code\cite{blaha}. The exchange-correlation functional has been treated by employing the recently developed PBEsol exchange-correlation functional \cite{perdew}. The muffin-tin radii were kept fixed to 2.0 Bohr (a$_{0}$) for all atoms, except for Sb, for which it was set equal to 2.2 a$_{0}$. The self-consistency  iteration was achieved by demanding the convergence of total energy/cell and charge/cell of the system to be less than 10$^{-4}$ Ry and 10$^{-3}$ electronic charge, respectively in every system. The transport properties calculations require the heavy k-point mesh in order to give more accurate results that is why the $50\times 50\times 50$ k-point mesh. To compute the transport properties of compounds, the BolzTrap code\cite{singh} under the constant relaxation time approximation for the charge carriers was employed. The lpfac was equal to 5, which represents the number of k-points per lattice point. The chemical potential in the calculation of transport properties was fixed to the middle of the gap in every case.

\section{Results and discussions}
Normally, the full Heusler compounds crystallize in L2$_{1}$ crystal structure, with space group Fm-3m. Therefore we have studied these compounds in the L2$_{1}$ phase. The crystal structure of Fe$_{2}$ScX, where X is atom of P, As or Sb is shown in Fig. 1. In this figure, Fe atoms are placed on the Wyckoff position 8c (1/4, 1/4, 1/4), Sc atoms are placed on the Wyckoff position 4a (0, 0, 0) and X atoms (i.e. P, As and Sb) are located at the Wyckoff position 4b (1/2, 1/2, 1/2). The cubic L2$_{1}$ structure consists of four interpenetrating fcc sublattices, two of which are equally occupied by Fe atoms. In these compounds the two Fe-site fcc sub-lattices, combine to form a simple cubic sub-lattice. The Sc and X atoms occupy alternatively, the centre of the simple cubic sub-lattice resulting in a CsCl-type superstructure.  

In order to obtain the equilibrium lattice parameters we have optimized the total energy as a function of volume of the cell. The obtained equilibrium lattice parameters for Fe$_{2}$ScP, Fe$_{2}$ScAs and Fe$_{2}$ScSb are 5.662, 5.811 and 6.058 {\AA}, respectively. The lattice parameter increases by about 7 \% on moving from Fe$_{2}$ScP to Fe$_{2}$ScSb. This is due to the reason that as one moves down the group in the periodic table, the atomic size increases and hence there is an increment in the lattice parameters. Using the equilibrium lattice parameters, firstly we have calculated the electronic properties of these compounds. The dispersion curves of Fe$_{2}$ScP, Fe$_{2}$ScAs and Fe$_{2}$ScSb are shown in the Figure 2. It is clear from the Fig. 2(a and b) that Fe$_{2}$ScP and Fe$_{2}$ScAs are semiconductors with indirect band gaps of about 0.3 and 0.09 eV, where the top of the valence band (VB) and bottom of the conduction band (CB) are at $\Gamma$ and X-point, respectively. In these materials the direct band gap is located at the $\Gamma$-point, which is $\sim$0.05 and 0.01 eV more than the value of indirect band gap of Fe$_{2}$ScP and Fe$_{2}$ScAs, respectively. Thus at higher temperature both direct and indirect transitions are expected to contribute to the transport properties of these materials. The dispersion curve of Fe$_{2}$ScSb is shown in Fig.2 (c). The CB edge at X-point is slightly lower in energy than the VB edge at $\Gamma$-point (not visible from the scale shown here) which make this compound semimetallic. The value of indirect band gap is found to decrease as one moves from the compound with P to the compound having Sb. This is because of decrease in electronegativity and the amount of orbital overlap, as moving down the group in the periodic table. In these materials, the Sb is the least electronegative element and form the weakest covalent bonding with Fe and thus band gap decreases from P to Sb atom. The top of the VB in all compounds is triply degenerate at $\Gamma$-point. Along $\Gamma$ to X-direction the degeneracy lifted and become doubly degenerate and non degenerate at X-point. Along L-direction the degeneracy is lifted to doubly degenerate and non degenerate band. Also along $\Gamma$ to X-direction there exist almost a flat conduction band in these compounds. Thus these compounds surely have large value of effective mass which suggests that these materials are good thermoelectric materials.  

The total density of states (TDOS) and partial density of states (PDOS) plots for Fe$_{2}$ScP, Fe$_{2}$ScAs and Fe$_{2}$ScSb are presented in the Figure 3. It is clear from the TDOS, shown in Fig. 3(a-c) that Fe$_{2}$ScP and Fe$_{2}$ScAs are semiconductors, while in Fe$_{2}$ScSb there is very small TDOS at the Fermi level (E$_{F}$) which is due to its semimetallic behavior of the compound as discussed above. In the VB region of these compounds two broad peaks with larger value of TDOS are found in the region between -4.0 to 0 eV and two broad peaks with smaller intensity exist between -7.0 to -4.0 eV. The maxima of these peaks move towards the higher energy as going from Fe$_{2}$ScP to Fe$_{2}$ScSb. The PDOS of Fe atoms of Fe$_{2}$ScP, Fe$_{2}$ScAs and Fe$_{2}$ScSb are shown in Fig. 3(d-f). These plots show that the top of the valence band has mainly \textit{t}$_{2g}$ character and bottom of the conduction band has \textit{e}$_{g}$ character. In the VB the \textit{t}$_{2g}$ with very small contribution of \textit{e}$_{g}$ states are extended in two different regions from -5 to -4 eV and -4 to 0 eV. In the region -5 to -4 eV the intensity of peaks are very small and \textit{t}$_{2g}$ states have more contribution than the \textit{e}$_{g}$ states. The contribution from \textit{t}$_{2g}$ (\textit{e}$_{g}$) states of Fe atom is 0.32 (0.14), 0.30 (0.13) and 0.26 (0.14) states/eV/atom in Fe$_{2}$ScP, Fe$_{2}$ScAs and Fe$_{2}$ScSb, respectively. On moving from Fe$_{2}$ScP to Fe$_{2}$ScSb these two regions become narrow and maximum of the peaks shifts slightly towards the higher energy. The PDOS of Sc atom of Fe$_{2}$ScP, Fe$_{2}$ScAs and Fe$_{2}$ScSb are presented in Fig. 3(g-i). It is evident from the graphs that there is small contribution from the \textit{d} orbitals of Sc atom than that of Fe atom. There are two broad peaks between -6 to -4 eV with smaller intensity and -3.5 to 0 eV having large intensity. The region nearer to the E$_{F}$ has more contribution from the \textit{t}$_{2g}$ and the region away from it is dominated by the \textit{e}$_{g}$ states. The maximum of the peaks shift towards higher energy from Fe$_{2}$ScP to Fe$_{2}$ScSb. The PDOS for \textit{p} orbitals of P, As and Sb atoms are shown in Fig. 3(j-l). There are two broad peaks in the region between -7 to -4 eV and as one move from P to Sb these peaks shift towards the higher energy. This trend is same as that is found in the atomic energy of \textit{p} orbitals of these atoms. The trend in the atomic energy of 3\textit{p}, 4\textit{p} and 5\textit{p} orbitals of P, As and Sb atom, respectively is -6.9 eV \textless -6.7 eV \textless -6.3 eV. In the PDOS plot of Fe and Sc atoms the very small peaks are appearing in the low energy region between -7 to -4 eV which are essentially due to the hybridization between the \textit{p} orbitals of P, As and Sb with the \textit{d} orbitals of Fe and Sc. This can be understood from the crystal structure shown in Fig. 1. This figure shows that the Fe atom is lying at the Wyckoff position 8c (1/4, 1/4, 1/4), thus there is more hybridization between the \textit{t}$_{2g}$ of Fe atom and \textit{p}-orbitals of X atom in comparison to the \textit{e}$_{g}$ orbitals. The Sc atoms are at origin and As atoms are lying at 4b (1/2, 1/2, 1/2). Thus in comparison to the \textit{t}$_{2g}$ states of Sc atom, more hybridization is taking place between the \textit{e}$_{g}$ orbitals of Sc and \textit{p}-orbital of X atom. This is evident from the PDOS plot for Fe and Sc atoms, where small peaks are appearing in the low energy region between -7 to -4 eV with dominating \textit{t}$_{2g}$ states in Fe and \textit{e}$_{g}$ states in Sc atom.

The presence of semiconducting ground state and flat CB, clearly suggest that they have potential to be used as a good thermoelectric materials. Therefore we have also study the thermoelectric properties of these materials. The computed transport coefficients of Fe$_{2}$ScP, Fe$_{2}$ScAs and Fe$_{2}$ScSb for chemical potential ($\mu$) ranging from -0.6 to 0.6 eV at different temperatures are shown in the Figure 4. The dashed line at $\mu$ = 0 eV represents the chemical potential for pure compound and was fixed to the middle of the gap. The positive value of $\mu$ represents the electron doping, whereas the negative $\mu$ means the hole doping. The Seebeck coefficient of Fe$_{2}$ScP, Fe$_{2}$ScAs and Fe$_{2}$ScSb, respectively at 200, 300 and 400 K as a function of $\mu$ is presented in Fig. 4(a-c). It is clear from these plots that at $\mu$ = 0 i.e. for pure compounds, the maximum value of $S$ is -770, -386 and -192$\mu$V/K for Fe$_{2}$ScP, Fe$_{2}$ScAs and Fe$_{2}$ScSb, respectively. The large and negative $S$ means that these compounds are of \textit{n}-type. The $S$ is found to be negative because of the presence of flat CB along $\Gamma$ - X-direction as stated earlier. For Fe$_{2}$ScP and Fe$_{2}$ScAs in the selected range of $\mu$, two peaks with maximum value of $S$ are obtained, one at positive $\mu$ and another at negative $\mu$. In Fe$_{2}$ScP at +14 meV the maximum value of $S$ are $\sim$ -853 and -608 $\mu$V/K for 200 and 300 K, respectively, however for 400 K the maximum $S$ is $\sim$ -486 $\mu$V/K and is obtained comparatively at higher $\mu$ equal to +21 meV. The other maximum having $S$ equal to $\sim$ 673, 433 and 316 $\mu$V/K was obtained at $\mu$ = -50, -70 and -90 meV for 200, 300 and 400 K, respectively. It is also clear from these plots that, when the temperature increases the maximum value of $S$ decreases and shifts slightly towards the higher $\mu$ (in magnitude). Similar trend is found in Fe$_{2}$ScAs where the maximum $S$ is $\sim$ -390, -300 and -270 $\mu$V/K at 200, 300 and 400 K, respectively for positive $\mu$. In Fe$_{2}$ScSb one maximum peak with same $S$ equal to $\sim$ 70 $\mu$V/ for all temperatures is found at negative $\mu$, while other maximum peak is obtained exactly at $\mu$ = 0 eV having value equal to $\sim$ -185 $\mu$V/K for all temperatures. In Fe$_{2}$ScP and Fe$_{2}$ScAs, the large $S$ is found for the positive value of $\mu$, i.e for electrons doping. This suggests that the \textit{n}-type doping will provide better thermoelectric behavior than the \textit{p}-type doping in these compounds.  

Fig. 4(d-f) shows the plots of electronic conductivity ($\sigma/\tau$) versus $\mu$ at different temperature for Fe$_{2}$ScP, Fe$_{2}$ScAs and Fe$_{2}$ScSb, respectively. The electronic conductivity increases with $\mu$ at same rate for all values of temperature, as is evident from these plots. The conductivity of these compounds increases with doping because of large carrier concentration with increasing $\mu$. The $\sigma/\tau$ is more for the positive $\mu$ in comparison to the negative value of $\mu$, which means that electron doped compounds will have more conductivity than hole doped. However, as one move from P to Sb containing compound the conductivity increases and maximum is obtained at the lower $\mu$. In Fe$_{2}$ScP the conductivity is almost same $\sim$ $10^{14}(\Omega^{-1}m^{-1}s^{-1}$) in the region between -140 to +90 meV. The width of region with almost same conductivity decreases in Fe$_{2}$ScAs. The electronic thermal conductivity ($\kappa_e$/$\tau$) as a function of $\mu$ for different values of temperature is presented in Fig. 4(g-i) for Fe$_{2}$ScP, Fe$_{2}$ScAs and Fe$_{2}$ScSb, respectively. From these plots it is evident that $\kappa_e$/$\tau$ increases with $\mu$ for all temperatures. However, at positive $\mu$, this value is found to be more than at negative $\mu$. This means that the largest value of $\kappa_e$/$\tau$ will be for electron doped compounds. 

A good thermoelectric material is characterized by its power factor and in order to improve the electronic properties of a thermoelectric material one has to enhance its power factor ($PF$). Therefore we have also shown the $PF$ as a function of $\mu$ for Fe$_{2}$ScP, Fe$_{2}$ScAs and Fe$_{2}$ScSb at 200, 300 and 400 K temperature in Fig. 4(j-l). It is clear from these plots that for every temperature there are two peaks having maximum $PF$ in the given range of $\mu$ and highest $PF$ is obtained at 400 K. Also it is clear from these plots that the maximum of these peaks shifts slightly towards the higher $\mu$ when temperature is increased from 200 to 400 K. The maximum $PF$ at negative $\mu$ increases from $\sim$ 14 to 46, 11 to 32 and 12 to 30 ($10^{14}\mu Wcm^{-1}K^{-1}s^{-1}$) for Fe$_{2}$ScP, Fe$_{2}$ScAs and Fe$_{2}$ScSb, respectively when temperature is increased from 200 to 400 K. The other maximum peak corresponding to large $PF$ exists at positive $\mu$ for Fe$_{2}$ScP and Fe$_{2}$ScAs, whereas in Fe$_{2}$ScSb it exists at $\mu$ = 0. The values of $PF$ is almost same for compounds with P and As and increases from 32 to 64, whereas in Fe$_{2}$ScSb it increases from 30 to 60 ($10^{14}\mu Wcm^{-1}K^{-1}s^{-1}$) with increasing temperature from 200 to 400 K. Thus plots of $PF$ versus $\mu$ provide the qualitative information that the $PF$ of the electron doped compounds is expected to be more. The position of maximum peak also show qualitatively that  the maximum $PF$ can be obtained when these compounds are heavily doped. 

In order to know the exact doping range to obtain the maximum $PF$ we have doped these compounds in the doping range between 10$^{17}$ to 10$^{21}$ cm$^{-3}$. From the above discussion it is clear that the $PF$ for electron doping will be more, however we have doped these compounds with electrons as well as holes. The transport coefficients versus temperature plots at different doping range are presented in Fig. (5 and 6). The $S$ versus temperature plots for p-type doping are shown in Fig. 5(a-c) and it is clear that the maximum peak of $S$ shifts towards the higher value of temperature as the concentration is increased. In Fe$_{2}$ScP and Fe$_{2}$ScAs, the maximum peak of $S$ exists at lower temperature for the carrier concentration of 10$^{17}$ cm$^{-3}$, whereas in Fe$_{2}$ScSb the maximum $S$ is found at higher temperature for the doping of 10$^{21}$ cm$^{-3}$. The maximum value of $S$ is $\sim$ 570, 280 and 86 $\mu$V/K at 200, 150 and 950 K for Fe$_{2}$ScP, Fe$_{2}$ScAs and Fe$_{2}$ScSb, respectively. For \textit{n}-type doping (Fig. 5(d-f)) the maximum peak of $S$ exists at 150 K in Fe$_{2}$ScP and Fe$_{2}$ScAs for the carrier concentration of 10$^{17}$ and 10$^{19}$ cm$^{-3}$, respectively whereas in Fe$_{2}$ScSb maximum is at 500 K for the carrier concentration of 10$^{20}$ cm$^{-3}$. The maximum value of  $S$ is $\sim$ -880, -460 and -180 $\mu$V/K for Fe$_{2}$ScP, Fe$_{2}$ScAs and Fe$_{2}$ScSb, respectively. Also it is clear from these plots that at the highest carrier concentration the value of $S$ remains negative for the entire temperature range in all compounds. In Fe$_{2}$ScSb the value of S is positive for the carrier concentration 10$^{17}$ to 10$^{19}$ cm$^{-3}$ and is almost same for all temperature. The $S$ versus temperature plots shows that the maximum $S$ is obtained when these compounds are doped with electrons and this is same as conjectured above. Also the highest value of $S$ is found in Fe$_{2}$ScP at low temperature for the carrier concentration of 10$^{17}$ cm$^{-3}$.  

The $\sigma/\tau$ versus temperature plots are shown in Fig. 5(g-l). These plots show that with increasing temperature the conductivity also increases in all these compounds and is almost same for the carrier concentration 10$^{17}$ to 10$^{20}$ cm$^{-3}$. However, for \textit{p}-type doping the maximum value of $\sigma/\tau$ is obtained at low temperature with 10$^{21}$ cm$^{-3}$ carrier concentration in all these compounds and its value is 0.33, 0.4 and 0.39 (10$^{20} \Omega^{-1}m^{-1}s^{-1}$) at 150 K for Fe$_{2}$ScP, Fe$_{2}$ScAs and Fe$_{2}$ScSb, respectively. Also in \textit{n}-type doping the maximum conductivity corresponds to the 10$^{21}$ cm$^{-3}$ carrier concentration in all these compounds. Fe$_{2}$ScP and Fe$_{2}$ScAs both have same value of $\sigma/\tau$ ($\sim$0.15 $\times 10^{20}\Omega^{-1}m^{-1}s^{-1}$) at 150 K and for Fe$_{2}$ScP this value corresponds to maximum conductivity. The maximum value of $\sigma/\tau$ for Fe$_{2}$ScAs and Fe$_{2}$ScSb is $\sim$0.2 and 0.3 (10$^{20}\Omega^{-1}m^{-1}s^{-1}$), respectively at 1200 K. 

Fig. 6 represents the $\kappa_e$/$\tau$ versus temperature and $PF$ versus temperature plots at different carrier concentration. The trend in the value of $\kappa_e$/$\tau$ with temperature is more or less the same for \textit{p} and \textit{n}-type doping. The thermal conductivity found to increase with temperature and have maximum value at the highest temperature under consideration. Among these compounds the maximum thermal conductivity is found in Fe$_{2}$ScSb for \textit{p} and \textit{n}-type doping. In moving from Fe$_{2}$ScP to Fe$_{2}$ScSb, the value of $\kappa_e$/$\tau$ increases from $\sim$0.18 to 0.25 (10$^{16}W m^{-1}K^{-1}s^{-1}$) for the given doping range expect for 10$^{18}$ cm$^{-3}$. For carrier concentration 10$^{18}$ cm$^{-3}$ maximum thermal conductivity increases from 0.22 to 0.24 $\times 10^{16}(W m^{-1}K^{-1}s^{-1}$) for \textit{p}-type doping and from 0.19 to 0.27 $\times 10^{16}(W m^{-1}K^{-1}s^{-1}$) in \textit{n}-type doping when one move from Fe$_{2}$ScP to Fe$_{2}$ScSb. 

The $PF$ versus temperature plots are presented in Fig. 6(g-l). It is clear from these plots that, the $PF$ increases slowly with temperature for the carrier concentration 10$^{17}$ to 10$^{19}$ cm$^{-3}$. The value of highest $PF$ is $\sim$ 10 and 7 (10$^{14}\mu Wcm^{-1}K^{-1}s^{-1}$) for Fe$_{2}$ScP and Fe$_{2}$ScAs, respectively at 1200 K when these compounds are doped with holes or electrons. However in Fe$_{2}$ScSb, the maximum $PF$ is obtained in the temperature range between 400 - 600 K with value of $PF$ closer to $\sim$ 1.5 (10$^{14}\mu Wcm^{-1}K^{-1}s^{-1}$). With the carrier concentration of 10$^{20}$ cm$^{-3}$, \textit{p}-doped Fe$_{2}$ScP, Fe$_{2}$ScAs and Fe$_{2}$ScSb have the maximum $PF$ equal to $\sim$ 22, 10 and 5 $\times 10^{14}(\mu Wcm^{-1}K^{-1}s^{-1}$) for temperature range between 700 - 800 K, 600 - 1200 K and 350 - 550 K, respectively. However, at same carrier concentration the \textit{n}-type doped Fe$_{2}$ScP, Fe$_{2}$ScAs and Fe$_{2}$ScSb have the maximum $PF$ equal to 23, 16, 7 (10$^{14}\mu Wcm^{-1}K^{-1}s^{-1}$), respectively for temperature range between 400 - 500, 200 - 300 and 100 - 150 K. In all these compounds the highest $PF$ is obtained for the  carrier concentration of 10$^{21}$ cm$^{-3}$ which corresponds to heavy doping of these compounds and this is also as per expectation. In moving from Fe$_{2}$ScP to Fe$_{2}$ScSb, the $PF$ decreases from $\sim$ 90 - 24 (10$^{14}\mu Wcm^{-1}K^{-1}s^{-1}$) for \textit{p}-type doping and $\sim$85 - 53 (10$^{14}\mu Wcm^{-1}K^{-1}s^{-1}$) for \textit{n}-type doping. In \textit{p}-doped Fe$_{2}$ScP and Fe$_{2}$ScAs the maximum $PF$ is obtained for the temperature range 1050 - 1200 K, whereas in Fe$_{2}$ScSb it is obtained for temperature ranging from 900 - 1050 K. The maximum $PF$ in \textit{n}-doped Fe$_{2}$ScP, Fe$_{2}$ScAs and Fe$_{2}$ScSb is obtained at lower temperature, ranging from 600 - 900, 400 - 700 and 400 - 600 K, respectively. 

It is clear from above discussion that the highest $PF$ is found for the heavily doped compounds. \textit{p}-doped Fe$_{2}$ScP and \textit{n}-doped Fe$_{2}$ScP and Fe$_{2}$ScAs have $PF$ \textgreater 60 for the temperature ranging from 700 - 1200, 400 - 1000 and 400 - 650 K, respectively. Also the obtained value of $PF$ for Fe$_{2}$ScP and Fe$_{2}$ScAs is comparable with the Bi$_{2}$Te$_{3}$\cite{singh}. The figure of merit of this compound is $\sim$ 1 at room temperature and is known as the one of the best commercially used thermoelectric material. Thus it is expected that Fe$_{2}$ScP and Fe$_{2}$ScAs can also have high value of figure of merit like Bi$_{2}$Te$_{3}$ and can be the potential thermoelectric materials in the wide range of temperature. Therefore the above results, suggest that when these compounds are heavily doped with electrons, the maximum $PF$ is obtained at the broad range of temperature, which is very important for various applications. 

From the above discussion it is evident that electron doped compounds have potential to be used as a good thermoelectric materials. However to the best of our knowledge no body has synthesized these compounds. In order to know whether these compounds will be synthesized or not we have calculated the formation energy ($FE$) of these compounds by using the following formula
\begin{equation}
FE = E (compound) - [2E (Fe) + E (Sc) + E (X)]
\end{equation}
where E (compound) is the energy of compound and E (Fe), E (Sc) and E (X) are the energies of the respective elements per atom. X represent the atom of P, As and Sb. Here we have considered the body centre cubic, face centre cubic and orthorhombic structure of Fe, Sc and P atom, while crystal structure of As and Sb is hexagonal as given in Myncryst database\cite{database}. These energies are obtained computationally by minimizing the total energy as a function of volume of the unit cell corresponds to above mentioned structures. The $FE$ comes out to be -2.2, -1.6 and -1.4 eV/formula unit for Fe$_{2}$ScP, Fe$_{2}$ScAs and Fe$_{2}$ScSb, respectively which clearly suggest that these compounds can be synthesized in the laboratories.

\section{Conclusions}
In the present work we have reported the three new full-Heusler alloys Fe$_{2}$ScP, Fe$_{2}$ScAs and Fe$_{2}$ScSb, which may be synthesized in the laboratories and expected to work as a good thermoelectric materials. The electronic and transport properties of these compounds are studied by using the first principles calculations with the Boltzmann transport theory. Fe$_{2}$ScP and Fe$_{2}$ScAs are indirect band gap semiconductors, whereas Fe$_{2}$ScSb is semimetallic. The value of band gap in Fe$_{2}$ScP and Fe$_{2}$ScAs is found to be 0.3 and 0.09 eV, respectively. We have observed the flat CB in these compounds along $\Gamma$ to X-direction which is mainly responsible for the large value of Seebeck coefficients observed in these compounds. The power factor of \textit{n}-doped Fe$_{2}$ScP and Fe$_{2}$ScAs comes out to be \textgreater 60 above room temperature and this value is comparable with the Bi$_{2}$Te$_{3}$, which is well known thermoelectric material. Our work shows that these materials have potentials to be used as a good thermoelectric materials which can operate over a wide temperature range.

\section{Figure Captions}

Fig. 1: (Color online) Crystal structures of Fe$_{2}$ScX, (X = P, As or Sb).

Fig. 2: (Color online) Dispersion curves of (a) Fe$_{2}$ScP, (b) Fe$_{2}$ScAs and (c) Fe$_{2}$ScSb.

Fig. 3: (Color online) Total and partial density of states plots for Fe$_{2}$ScP, Fe$_{2}$ScAs and Fe$_{2}$ScSb. Shown are (a, b and c) the TDOS plots, (d, e and f) PDOS of Fe atoms, (g, h and i) PDOS of Sc atoms and (j, k and l) PDOS of P, As and Sb
atoms.

Fig. 4: (Color online) Variation of transport coefficients with $\mu$. (a, b and c) Seebeck coefficient with $\mu$, (d, e and f) Electrical conductivity with $\mu$, (g, h and i) Electronic thermal conductivity with $\mu$ and (j, k and l) power factor with $\mu$ for Fe$_{2}$ScP, Fe$_{2}$ScAs and Fe$_{2}$ScSb.

Fig. 5: (Color online) Seebeck coefficient versus temperature plots (a, b and c) \textit{p}-doped and (d, e and f) \textit{n}-doped Fe$_{2}$ScP, Fe$_{2}$ScAs and Fe$_{2}$ScSb. And electronic conductivity versus temperature plots (g, h and i) \textit{p}-doped and (j, k and l) \textit{n}-doped Fe$_{2}$ScP, Fe$_{2}$ScAs and Fe$_{2}$ScSb.

Fig. 6: (Color online) Electronic thermal conductivity versus temperature plots (a, b and c) \textit{p}-doped and (d, e and f) \textit{n}-doped Fe$_{2}$ScP, Fe$_{2}$ScAs and Fe$_{2}$ScSb. And power factor versus temperature plots (g, h and i) \textit{p}-doped and (j, k and l) \textit{n}-doped Fe$_{2}$ScP, Fe$_{2}$ScAs and Fe$_{2}$ScSb.

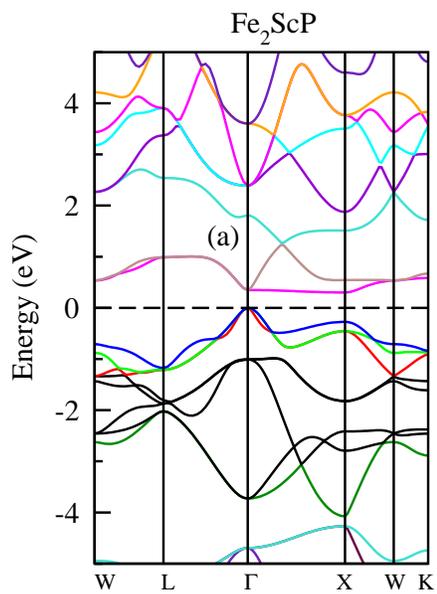
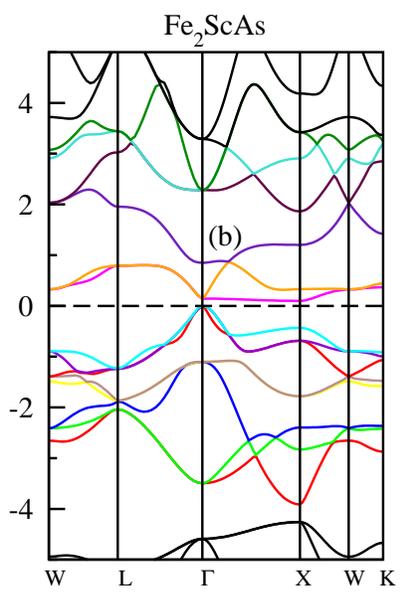
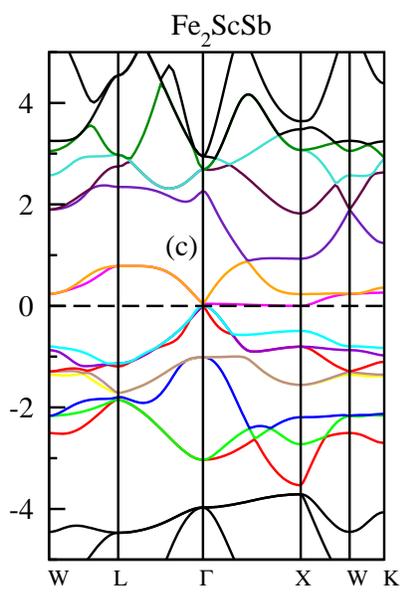
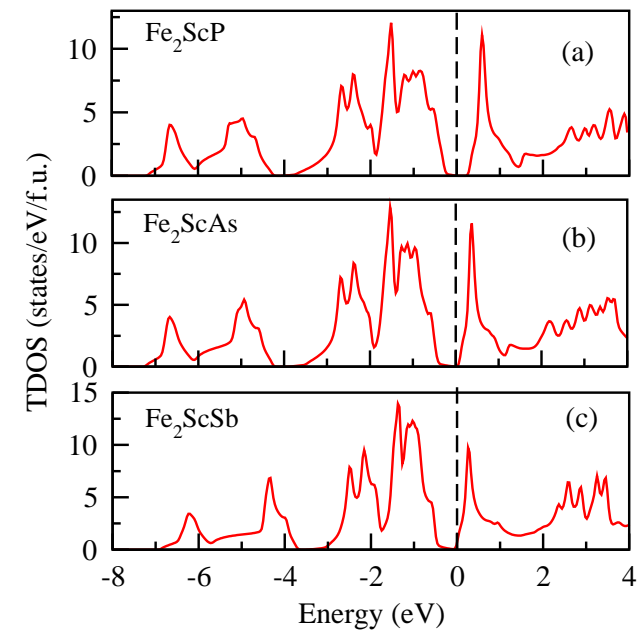
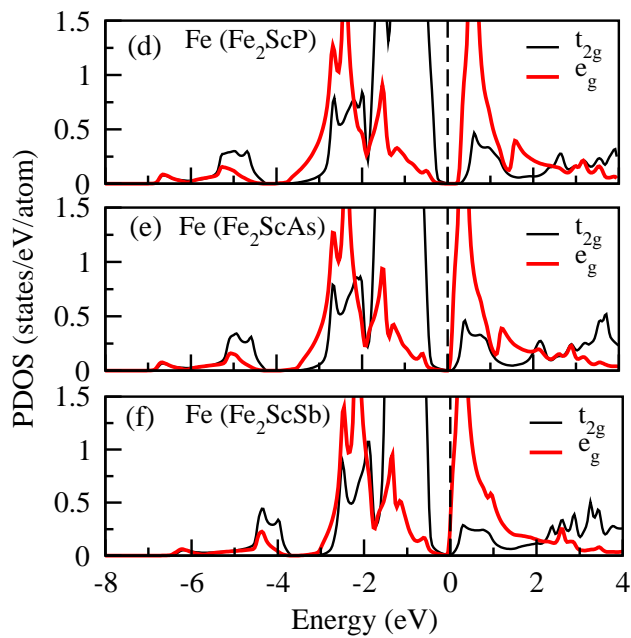
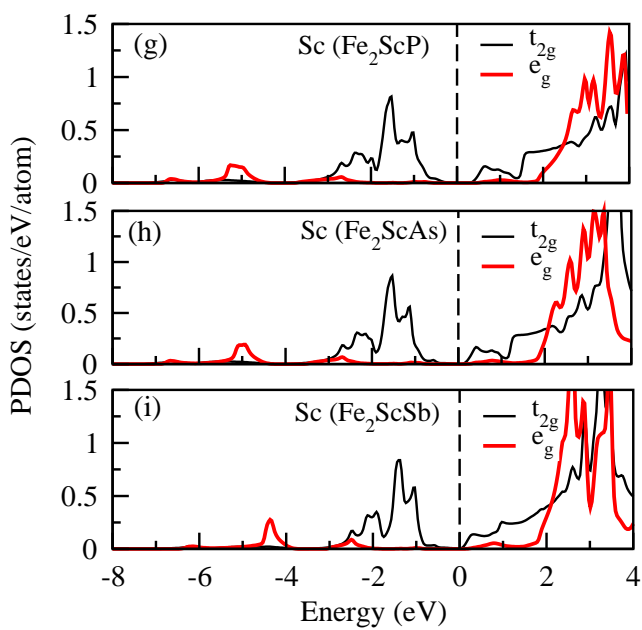
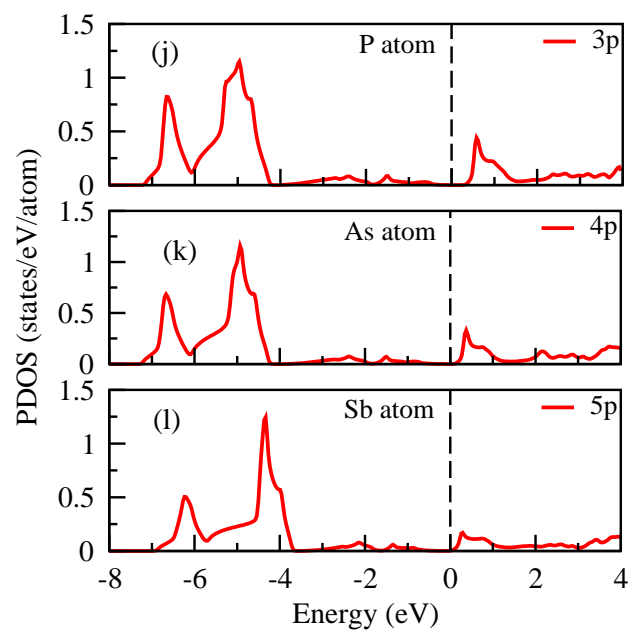
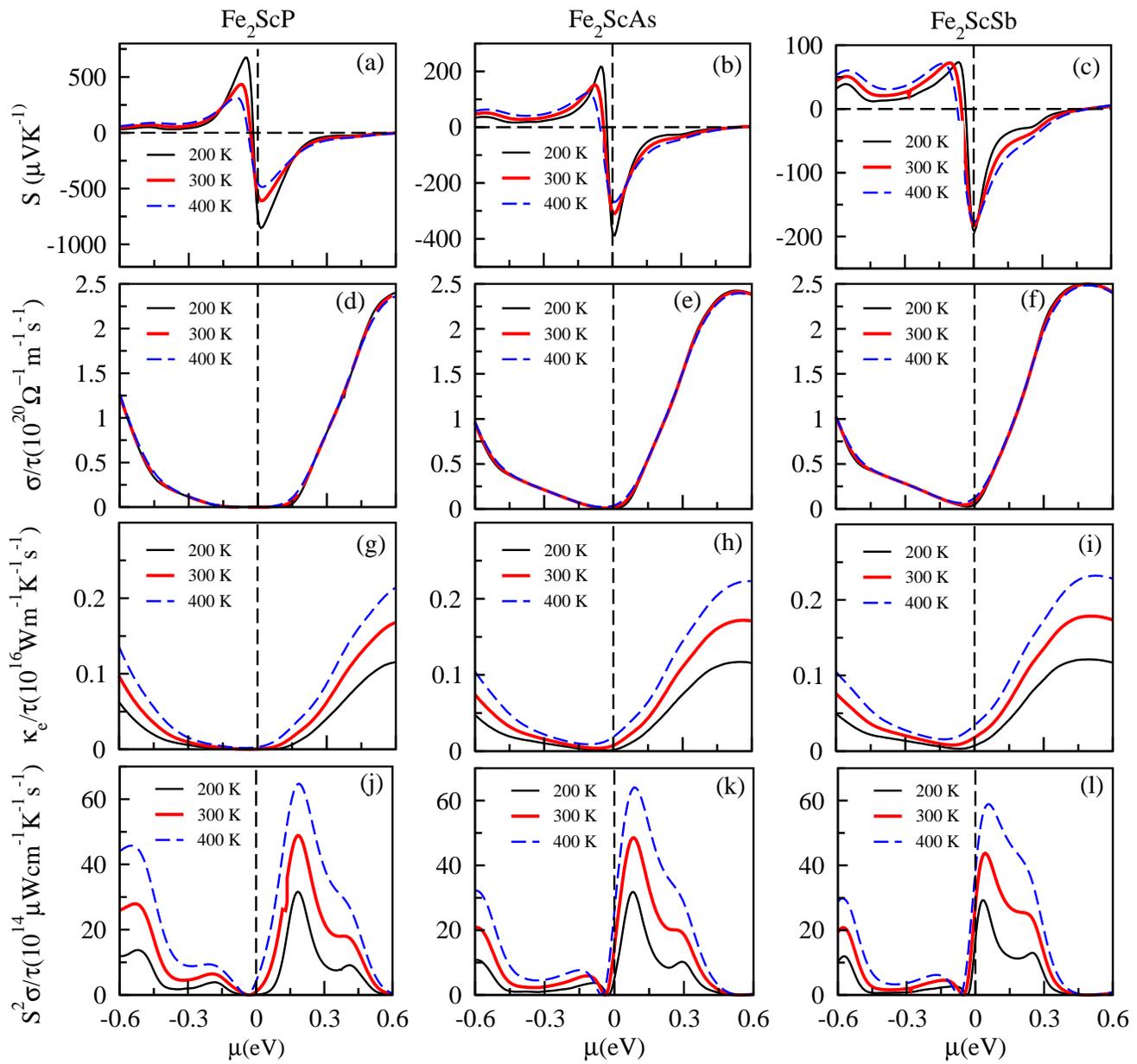
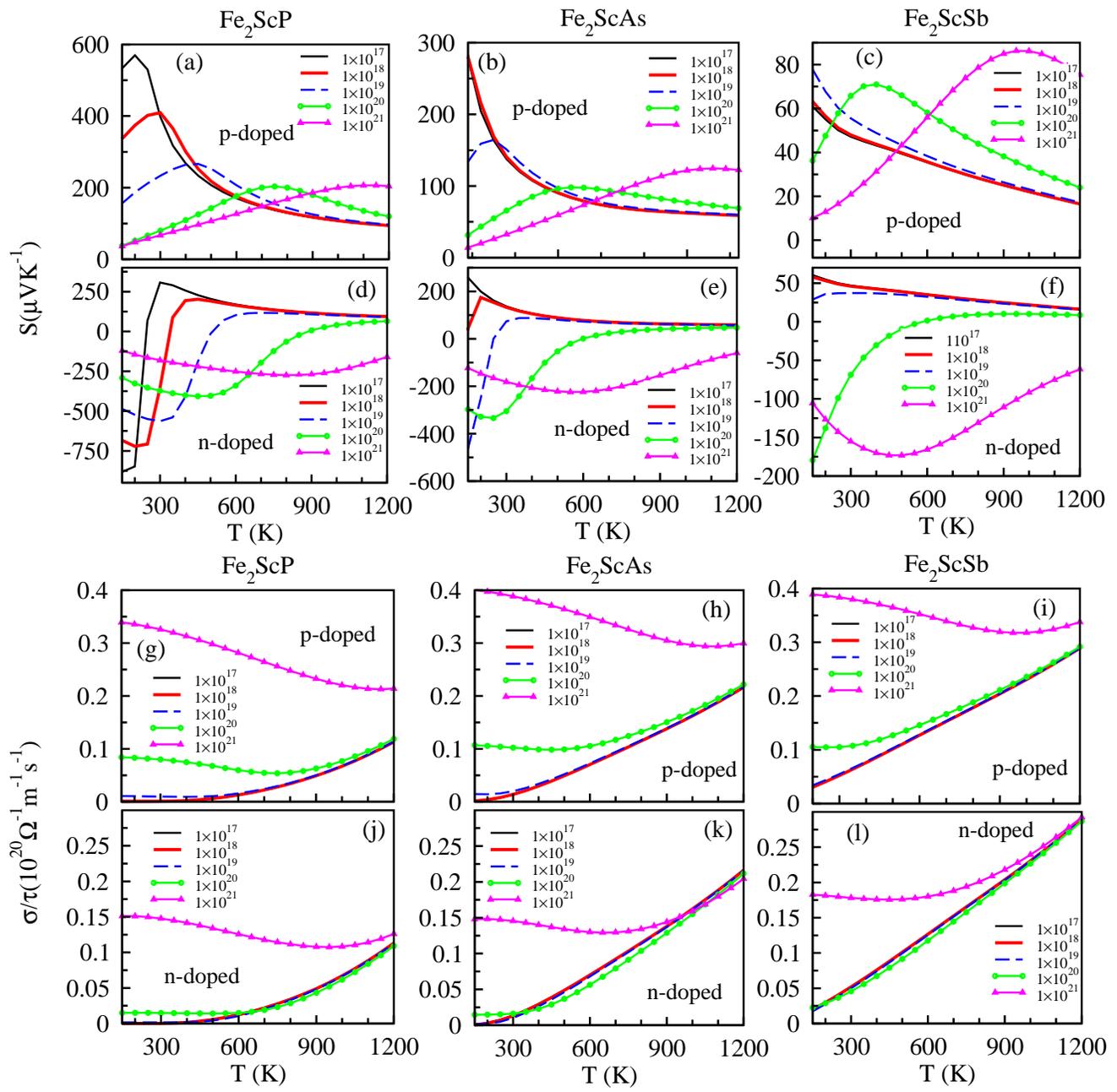
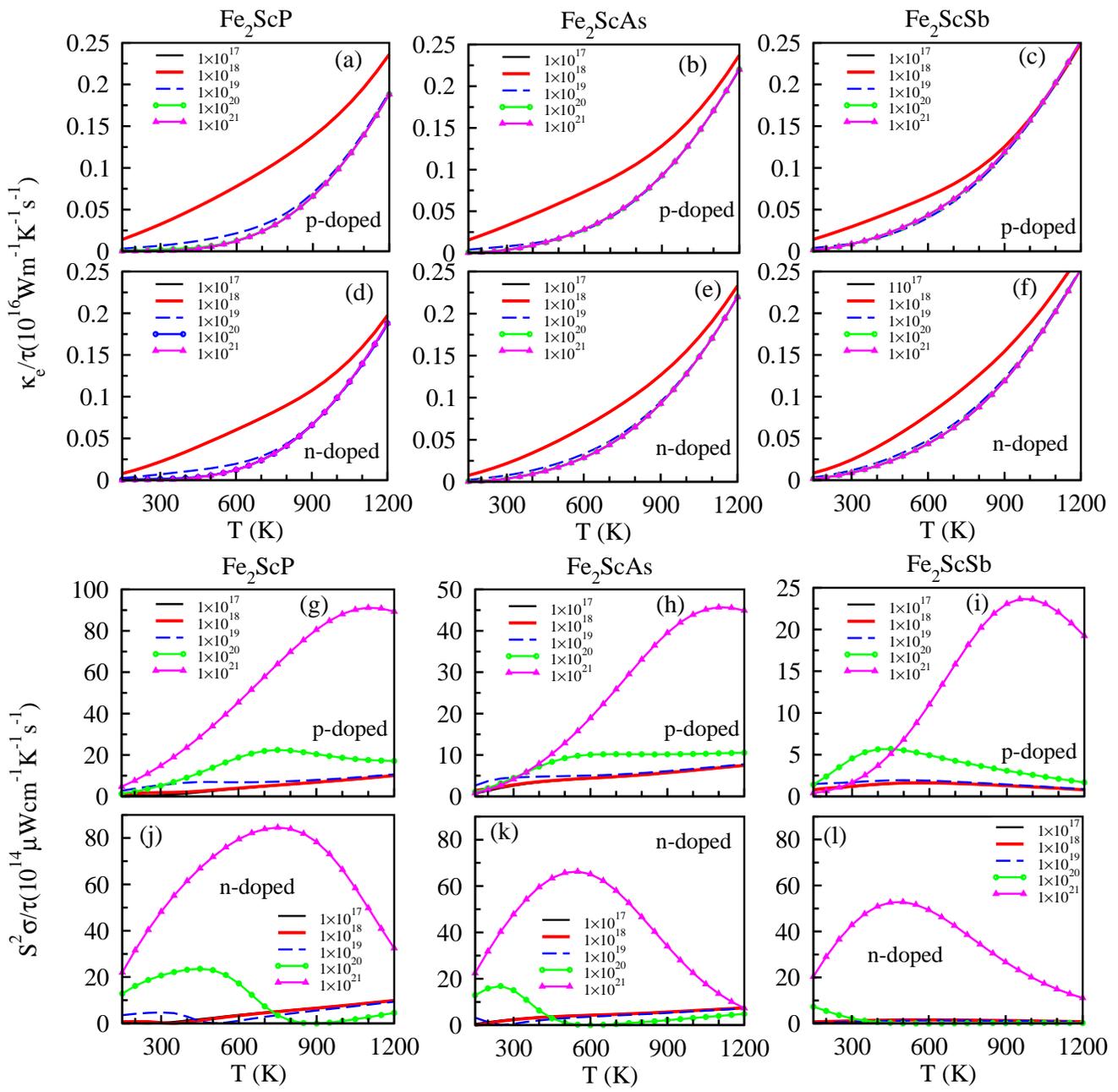
\end{document}